\documentclass[a4paper,11pt]{article}
\pdfoutput=1 

\usepackage{jcappub} 

\usepackage[T1]{fontenc} 

\usepackage{microtype}
\usepackage{enumitem}

\usepackage{physics}

\usepackage{natbib}

\title{\boldmath Asymmetry of the CMB  map: local and global anomalies}

\author{James Creswell}
\author{and Pavel Naselsky}
\affiliation{Niels Bohr Institute, University of Copenhagen, Blegdamsvej 17, DK-2100 Copenhagen, Denmark}

\emailAdd{james.creswell@nbi.ku.dk}
\emailAdd{naselsky@nbi.dk}

\abstract{We investigate the sources of parity asymmetry in the CMB temperature maps using a pixel domain approach. We demonstrate that this anomaly is mainly associated with the presence of two pairs of high asymmetry regions. The first pair of peaks with Galactic coordinates $(l, b) = (212^\circ, -21^\circ)$ and $(32^\circ, 21^\circ)$ is associated with the Northern Galactic Spur and the direction of the dipole modulation of the power spectrum of the CMB anisotropy. The other pair ($(l, b)=(332^\circ, -8^\circ)$ and $(152^\circ, 8^\circ)$) is located within the Galactic plane (the Galactic Cold Spot and its antipodal partner). Similar asymmetric peaks, but with smaller amplitudes, belong to the WMAP/Planck Cold Spot and its partner in the Northern Galactic Spur. These local anomalies increase the odd-multipole power to a level consistent with Gaussian simulations. In contrast, the deficit of symmetric peaks is accompanied by a deficit in the even-multipole power and is the source of the parity asymmetry of the CMB temperature maps at the level of about 3 sigma. We also evaluate the influence of the quadrupole, which is another source of the even-multipole deficit. If the low quadrupole is an intrinsic feature of the theoretical model, it will reduce the significance of the parity asymmetry to around the 2 sigma level. We also investigate the relationship between the asymmetry of the power spectrum and the level of the parity asymmetry in the framework of a model with dipole modulation of a statistically uniform Gaussian signal. We show that these two anomalies are innately linked to each other.}

\begin{document}
\maketitle
\flushbottom

\section{Introduction}
\label{sec:intro}

The concept of symmetry and its breaking constitute the foundation of modern particle physics.
It is clear that it is also important for the study
of the statistical properties of CMB \cite{IS2015,IS2018}.
One of the most statistically significant anomalies of the CMB temperature in the low multipole domain ($\ell \le 50$) of the power spectrum $C_\ell$ is the hemispherical power asymmetry \cite{Eriksen_2004,Hansen_2009} and the point parity asymmetry \cite{Kim_2010}.

Non-Gaussianity of the CMB at low multipoles has been a longstanding area of interest.
In \cite{Copi_2004}, Gaussianity and isotropy for $2 \leq \ell \leq 8$ was rejected with significance $p \approx 0.01$ in the preliminary WMAP data, and in \cite{Schwarz_2004} the alignment of the $\ell = 2$ and $\ell = 3$ modes was detected with significance $p \approx 0.001$.
These and other low-$\ell$ anomalies were widely investigated in the context of the WMAP data, and several theoretical and systematic explanations were discussed (see e.g.\ \cite{de_Oliveira_Costa_2004,PhysRevD.72.101302,PhysRevLett.95.071301}). 
The hemispherical power asymmetry was further explored and confirmed with high significance using different techniques in \cite{Eriksen_2004b,Hansen_2004,Eriksen_2007,Hoftuft_2009}.
Parity asymmetry, the apparent excess of power in the odd-$\ell$ multipoles compared to the even-$\ell$ multipoles was revisited in \cite{Aluri_2011,Hansen_2011,Kim_20102,Gruppuso_2010,Kim_2012} using the WMAP 5-year and 7-year data releases. 
In these works, which focused on estimators of the power spectrum in the multipole range $2 \leq \ell \leq 20$, the odd-parity excess was detected with significance $p = 0.004$, and in \cite{Kim_2012} this result was linked with the correlation function on large angular scales.
More recently, the problem of the alignment of the quadrupole, octopole, and other odd-$\ell$ multipoles has been confirmed in the Planck data with significance of 2 to 3 sigma using various estimators \cite{Copi_2015, Aluri_2017}, and parity asymmetry and other low-$\ell$ anomalies have also been detected and investigated using the Planck data \cite{Akrami_2014,Gruppuso:2017nap,IS2018,Shaikh_2019}.

The goal of our paper is to perform a new pixel-domain analysis and argue that:
\begin{enumerate}[label=\alph*),noitemsep,topsep=0pt]
    \item the hemispherical power asymmetry and the parity asymmetry have a common origin;
    \item they are associated with asymmetric ditribution of high amplitude peaks;
    \item the parity asymmetry clearly indicates the deficit of negative peaks;
    \item  the sources of asymmetry are only partially associated with the Galactic plane.
\end{enumerate}

Unlike standard analysis of asymmetries in the $C_\ell$ domain, we perform our investigation purely in the pixel domain. 
This allows us to identify zones of the sky that make the main contribution to
anomalies. 
Smoothing maps with a Gaussian filter effectively removes some of the multipoles from the analysis:
the smoothing angle $\Theta$, which is the FWHM of the Gaussian filter, is connected with the cerresponding multipole $\ell$ via the approximate relationship $\Theta \simeq 100^\circ / \ell$.

The comparison between pixel and multipole approaches cannot be done
without implementation of different masks.
Unless otherwise stated, our standard mask in use is a ring masking Galactic latitudes $-10^\circ < b < 10^\circ$.
Such a mask is quite similar to the Planck confidence mask, however, it has the important property that a pixel $\vb{n}$ is masked if and only if the corresponding opposite pixel $-\vb{n}$ is masked, which simplifies the calculation of the parity asymmetry pixel domain estimators below.

The outline of our paper is the following: first, in Section \ref{sec:2}, we review the decomposition of the sky into symmetric and asymmetric components, and define the pixel-domain $Z(\mathbf{n})$ asymmetry operator in use for this work. In Sections \ref{sec:3} and \ref{sec:4} we study the morphology and the statistics of this estimator, which reveals the sources of the parity asymmetry. In Section \ref{sec:5} we compare to Gaussian simulations, especially focusing on the contribution of the quadrupole deficit to the overall parity asymmetry, and the effect of changing cosmological parameters. In Section \ref{sec:6} we show how this is related to the power asymmetry and a brief conclusion is in Section \ref{sec:7}.

\section{Mathematical basis}
\label{sec:2}

In the CMB temperature map $T(\vb{n})$, for each pixel $\vb{n}$, one can denote the symmetric $S(\vb{n})$ and asymmetric $A(\vb{n})$ parts:
\begin{subequations} \label{eq1}
\begin{align}
    S(\vb{n}) &= \frac{T(\vb{n}) + T(-\vb{n})}{2}, \\
    A(\vb{n}) &= \frac{T(\vb{n}) - T(-\vb{n})}{2},
\end{align}
\end{subequations}
where we have $T(\vb{n})=S(\vb{n})+A(\vb{n})$.
In the multipole domain, these components can be decomposed through the spherical harmonics with the corresponding projection operators
$\gamma_s= \cos^2\qty(\frac{\pi \ell}{2})$ and $\gamma_a=\sin^2\qty(\frac{\pi \ell}{2})$ as follows:
\begin{subequations}
\begin{align}
S(\vb{n}) &=\sum_{\ell=2}^{\ell_\mathrm{max}} \sum_{m=-\ell}^\ell a_{\ell m} Y_{\ell m}(\vb{n})\gamma_s,\qquad\\
A(\vb{n}) &=\sum_{\ell=2}^{\ell_\mathrm{max}} \sum_{m=-\ell}^\ell a_{\ell m} Y_{\ell m}(\vb{n})\gamma_a,\\
T(\vb{n}) &=\sum_{\ell=2}^{\ell_\mathrm{max}} \sum_{m=-\ell}^\ell a_{\ell m} Y_{\ell m}(\vb{n}),
\label{eq2}
\end{align}
\end{subequations}
where $a_{\ell m}$ are the coefficients of decomposition.

For estimation of the degree of asymmetry of the CMB map, we will use the following function:
\begin{align}
Z(\vb{n})=T(\vb{n})T(-\vb{n})=S^2(\vb{n}) - A^2(\vb{n}).
\end{align}
If $Z(\vb{n})>0$ for the point $\vb{n}$ we have dominance of the symmetric part, and vice versa, if $Z(\vb{n})<0$ the asymmetric part is greater than the symmetric part. 

Averaging over all pixels of the map leads to the following result:
\begin{equation}
\langle Z(\vb{n})\rangle=\sum_\ell(2l+1)C_{\ell}\cos(\pi\ell),
\label{average}
\end{equation}
where $C_{\ell}$ corresponds to the actual realization of the theoretical power spectrum.  
Due to the properties of spherical harmonics, any even-$\ell$ mode has $Z(\vb{n}) \geq 0$, and any odd-$\ell$ mode has $Z(\vb{n}) \leq 0$.
Consequently, the full sky mean of $Z(\vb{n})$ is not expected to be $0$.
Instead, it has some positive bias (in average over statistical ensemble of realizations) because the leading-order term is the quadrupole, which always has $Z(\vb{n}) > 0$.

\section{\boldmath Visualization of \texorpdfstring{$Z(\vb{n})$}{Z(n)}}
\label{sec:3}

\begin{figure}[t]
    \centering
    \includegraphics{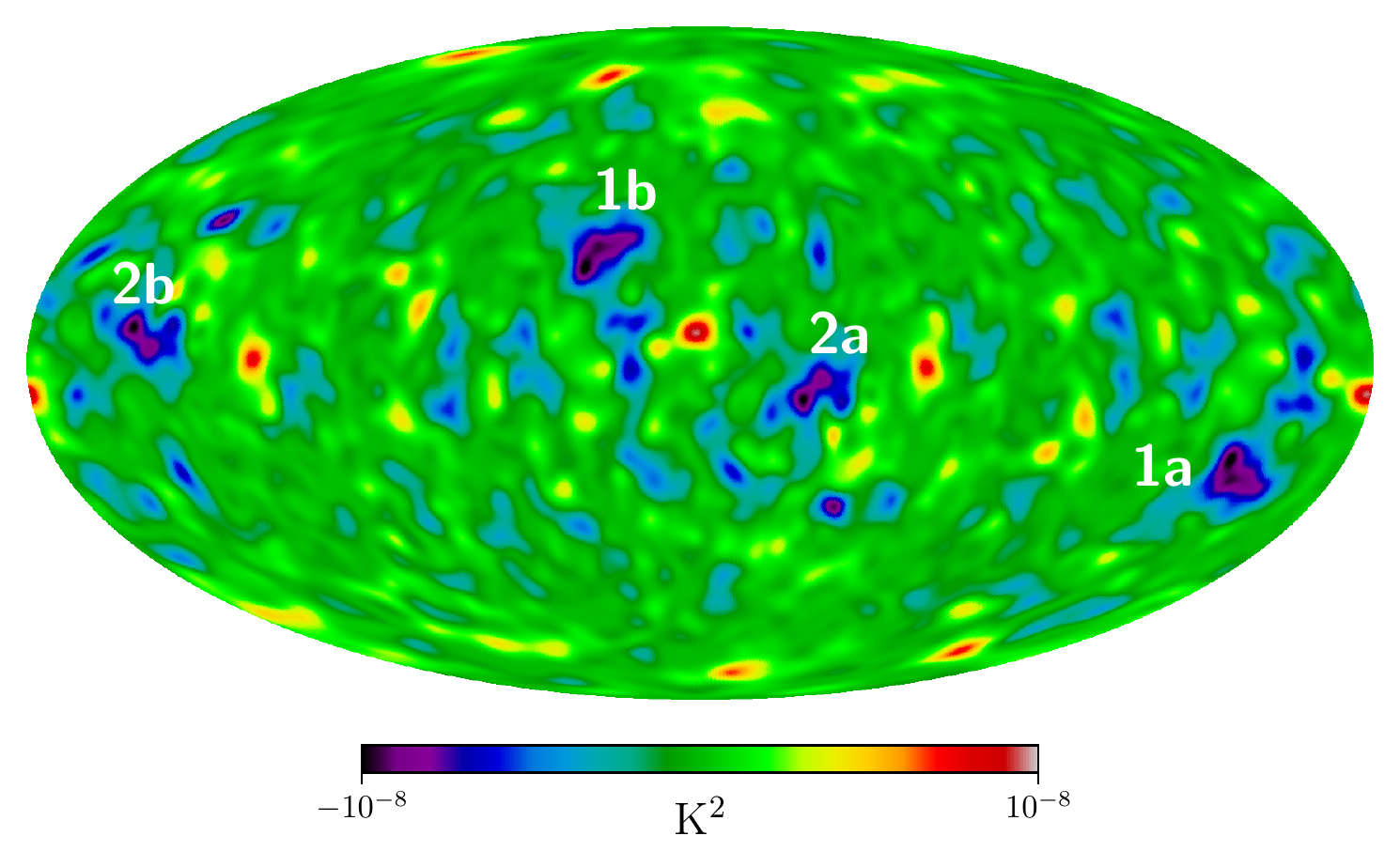}
    \caption{Morphology of $Z(\vb{n})$ map of Planck 2018 SMICA with $5^\circ$ smoothing. Note that the peaks 1a and 1b correspond to inversion $\vb{n}\rightarrow -\vb{n}$, as for 2a and 2b as well.}
    \label{fig:map}
\end{figure}

By definition $Z(\vb{n})$ is a local estimator, and we can plot its map on the sky. 
We show $Z(\vb{n})$ without masking for the Planck 2018 SMICA map \cite{CS2018} with 5 degrees smoothing in figure~\ref{fig:map}.
Note that by construction this map is symmetric, $Z(\vb{n}) = Z(-\vb{n})$.
The most interesting observation from figure~\ref{fig:map} is the two pairs of very strong high negative peaks of $Z(\vb{n})$ (labelled 1a/1b and 2a/2b) and about twenty negative peaks with smaller amplitudes, mainly localized within the area $|b|\le 30^\circ$ in Galactic coordinates.
The strongest peaks have the following coordinates $(l, b)$:
\begin{alignat*}{2}
    &\mathrm{1a}: (212^\circ, -21^\circ), \qquad &&\mathrm{1b}: (32^\circ, 21^\circ)\\
    &\mathrm{2a}: (332^\circ, -8^\circ), \qquad &&\mathrm{2b}: (152^\circ, 8^\circ)
\end{alignat*}
There are about nine pairs of high amplitude positive peaks, as it is seen from figure~\ref{fig:map},  localized in the same $|b|\le 30^\circ$ belt, and about six highest peaks distant from this area near the poles.

It is noteworthy that the coordinates of the 1a peak are practically the same as the direction found for the power asymmetry in the model of dipole modulation, e.g.~\cite{Hoftuft_2009} reports $(l,b) \approx (224^\circ,-22^\circ)\pm 24^\circ$, and \cite{IS2018} adopts $(l,b) = (221^\circ,-20^\circ)$ for the dipole modulation direction. The peak 1b is nothing but the symmetric
partner of the peak 1a. The peak 2a  coincides with the position of the Galactic Cold Spot, and 2b is its antipodal partner. These directions will be excluded from the analysis of the power asymmetry due to proximity to the Galactic plane where they are masked out. In the forthcoming sections
we will show that these anomalies, the parity and power asymmetries, could have the same origin.

\section{\boldmath Statistical properties of \texorpdfstring{$Z(\vb{n})$}{Z(n)}}
\label{sec:4}

\subsection{Gaussian theory}

\begin{figure}[t]
    \centering
    \includegraphics[width=0.7\textwidth]{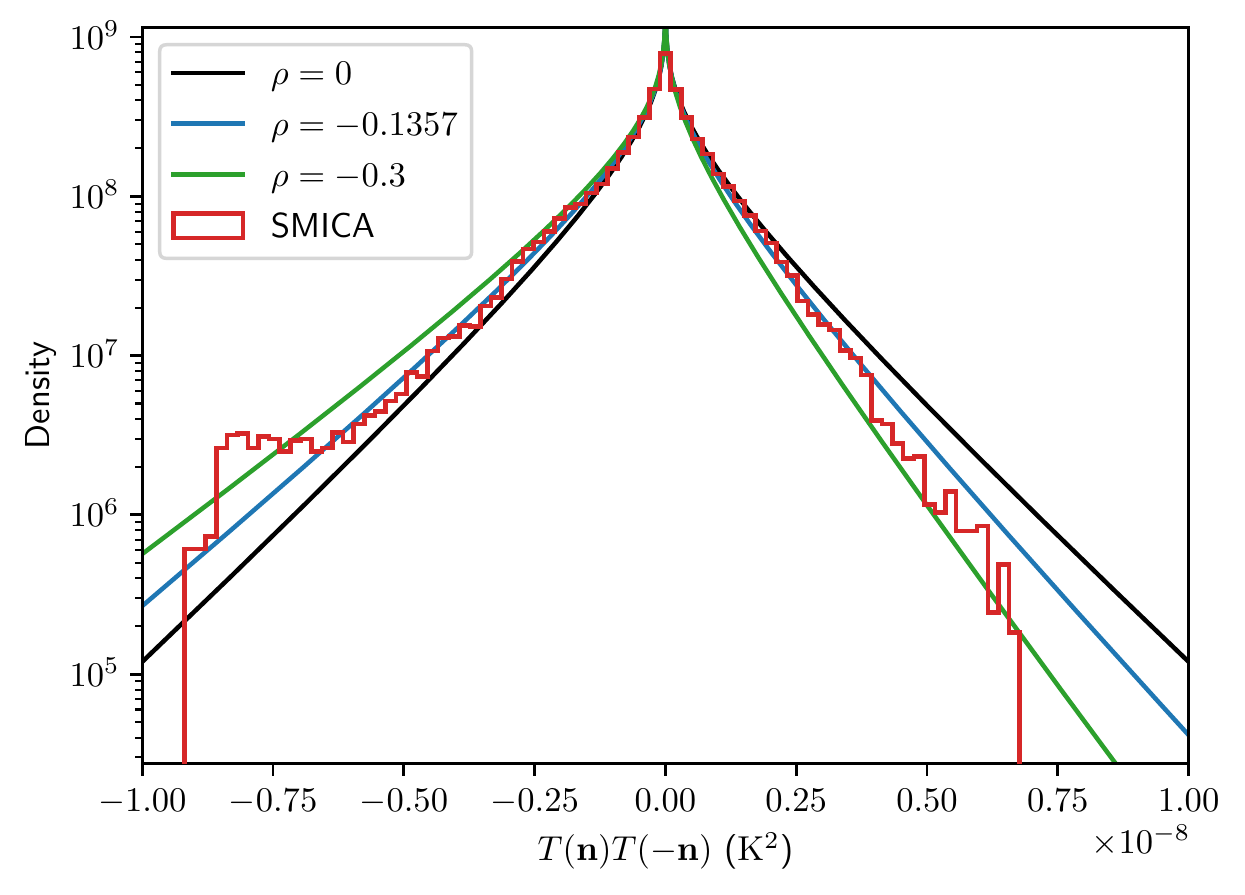}
    \caption{Theoretical $Z(\vb{n})$ distribution functions according to the Gaussian model, for $\rho = 0$ (black), $\rho = -0.1357$ (blue), and $\rho=-0.3$ (green). This is compared to the actual calculated distribution function of SMICA, shown in red.}
    \label{fig:Z}
\end{figure}

Although we will work with simulated sky maps for estimation of the significance level of the anomalies, it is also valuable to have a basic understanding of the statistical properties of function $Z(\vb{n})$. 
Suppose $T(\vb{n})$ is a realisation of  statistically isotropic Gaussian field.
Then, $T(-\vb{n})$ is also Gaussian-distributed.
However, the two quantities $T(\vb{n})$ and $T(-\vb{n})$ are not independent random variables due to correlations in the pixel domain.
The precise details of this correlation are determined by the power spectrum and the smoothing angle $\Theta$.

For our purposes, it is sufficient to consider the Pearson cross-correlation coefficient of $T(\vb{n})$ and $T(-\vb{n})$:
\begin{eqnarray}
\label{eq:rho}
\rho = \mathrm{Corr}(T(\vb{n}), T(-\vb{n})) = \frac{\displaystyle \int T(\vb{n}) T(-\vb{n}) \, d\vb{n}  }{\displaystyle \int T(\vb{n})^2 \, d\vb{n}} = \frac{ \sum_{\ell}(-1)^{\ell} \sum_m |a_{\ell m}|^2 }{\sum_{ \ell} \sum_m |a_{\ell m}|^2 }. 
\end{eqnarray}
We have taken that the mean is subtracted, $\int T(\vb{n}) \, d\vb{n} = 0$. The integrals are calculated as sums over all available pixels.

For given $\rho$, the distribution function of $Z(\vb{n})$ has a form \cite{NADARAJAH2016201,gaunt}:
\begin{equation}
    \label{eq:P(Z)}
    \mathcal{P}(Z') = \frac{1}{\pi \sqrt{1 - \rho^2}} \exp(\frac{\rho Z'}{1 - \rho^2}) K_0 \qty(\frac{|Z'|}{1 - \rho^2}), 
\end{equation}
where $K_0$ is the 0-th order modified Bessel function of the second kind
and $Z' = Z / \mathrm{var}(T)$ is the rescaled $Z$ to unit variance.
In figure~\ref{fig:Z}, function $\mathcal{P}(Z)$ from equation~\eqref{eq:P(Z)} is plotted for three sample values of $\rho$.
When $\rho = 0$, the distribution is symmetric. 
When $\rho < 0$, the distribution skews to the left and  corresponds to an excess of odd-$\ell$ power.
The value of $\rho$ estimated from the SMICA data is $\rho_\mathrm{SMICA} = -0.1357$. 
Therefore, the simple cross-correlation estimator indicates an excess of odd-$\ell$ power, or a deficit of even-$\ell$ power.

However, as seen in figure~\ref{fig:Z}, the SMICA distribution function deviates quite strongly from the model with $\rho_\mathrm{SMICA} = -0.1357$: on the left, there is a strong bump around $Z\sim -7 \times 10^{-8}$, and on the right above $Z > 3 \times 10^{-8}$, there is a correspondingly smaller observed density in the histogram. These features will be investigated more carefully using $\Lambda$CDM simulations.

\subsection{Comparison to simulations}

Figure~\ref{fig:map} reveals one of the advantages of the local estimator: we can directly identity the parts of the sky that are the main sources of asymmetry.
For example, the purple and black zones mark pixels $\vb{n}$ in which $T(\vb{n})$ and $T(-\vb{n})$ have simultaneously large amplitudes and different signs; the red and white zones are pixels in which $T(\vb{n})$ and $T(-\vb{n})$ both have large amplitudes and they also have the same signs.
To characterize the occurrence of such features in a general way, we investigate to the distribution function of $Z(\vb{n})$, which in practice means calculating its histogram over the observed sky.

\begin{figure}[!t]
    \centering
    \includegraphics[width=0.49\textwidth]{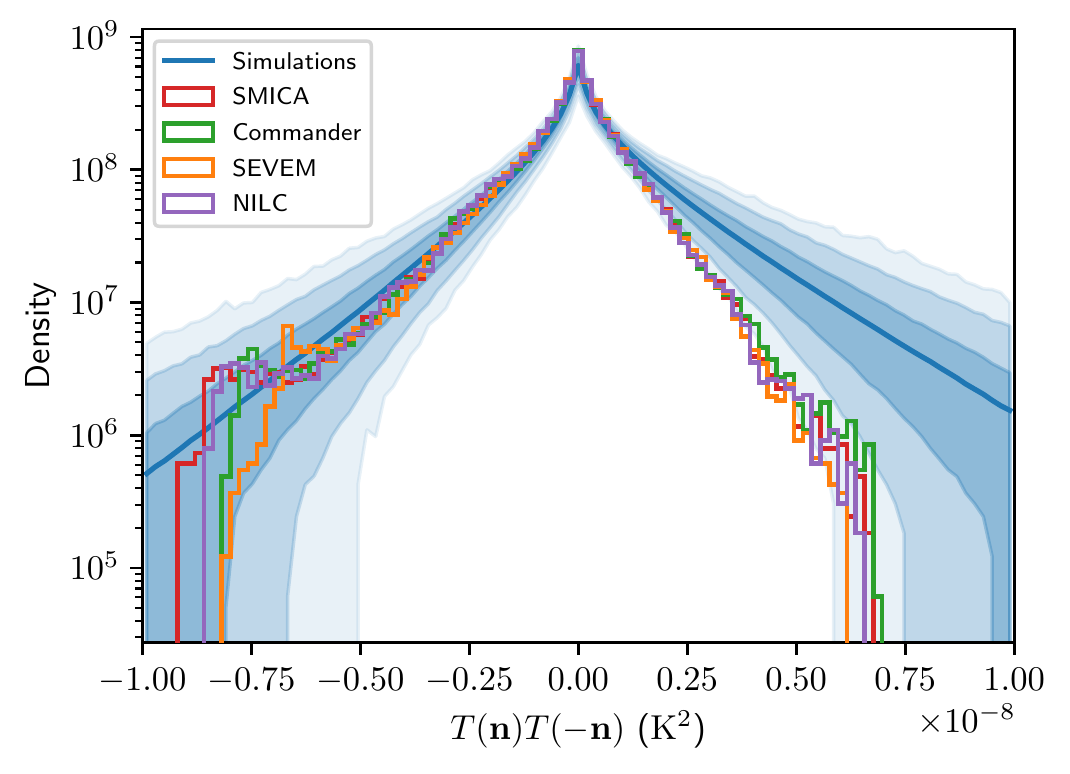}
    \includegraphics[width=0.49\textwidth]{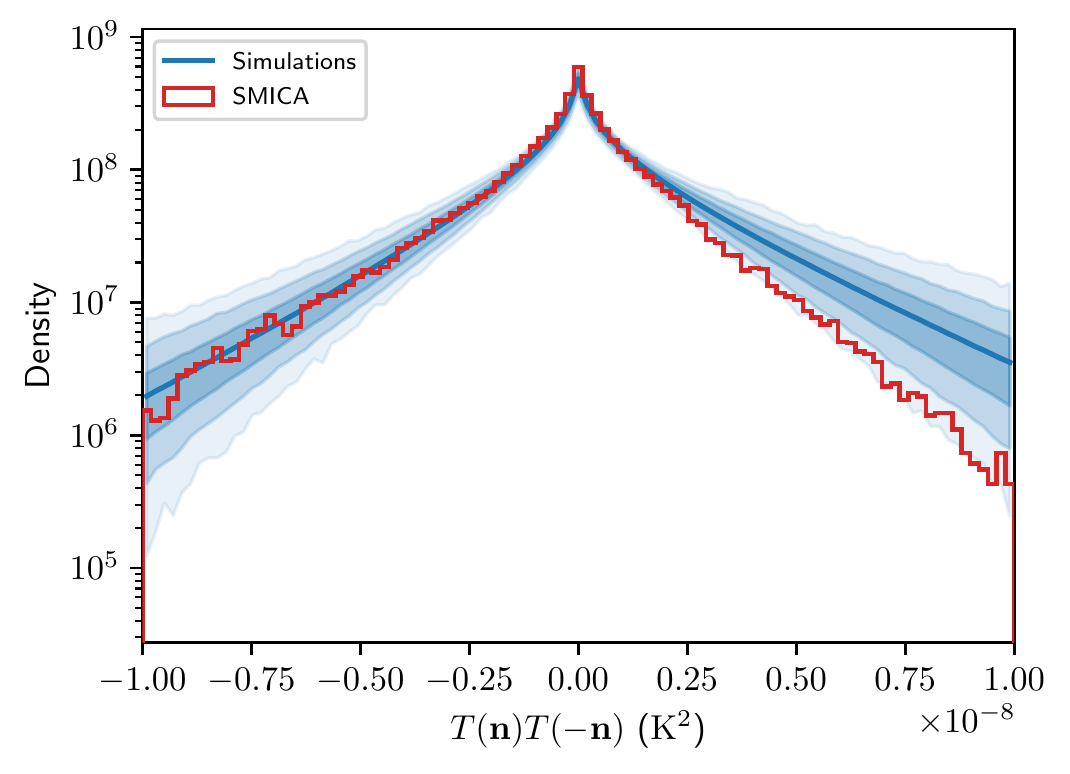}
    \caption{Left panel: $Z(\vb{n})$ distribution functions from CAMB simulations (blue) and actual distribution function from SMICA (red) and the other Planck maps. The smoothing angle is $\Theta = 5^\circ$. The 68\%, 95\%, and 99.7\% uncertainty regions of the simulations are shown in the three shades of blue. On the right side of the distribution, we have a departure of the distribution function at around or greater than the $3\sigma$ level. Right panel: the same for SMICA with smoothing angle $\Theta = 2.5^\circ$.}
    \label{fig:hists}
\end{figure}

\begin{figure}[!tbh]
    \centering
    \includegraphics[width=0.45\textwidth]{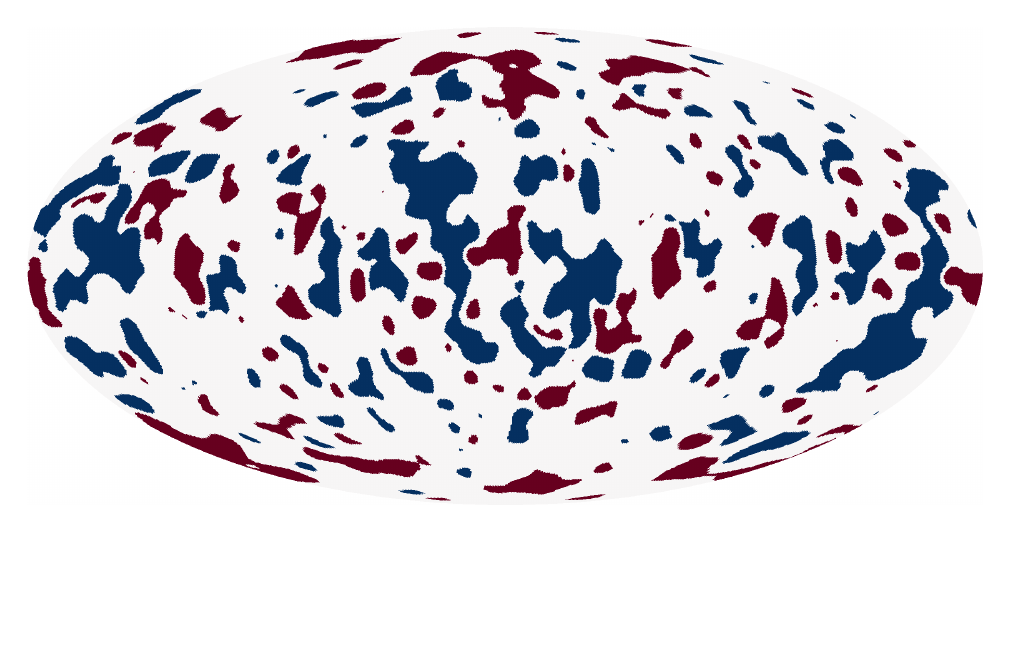}
    \includegraphics[width=0.45\textwidth]{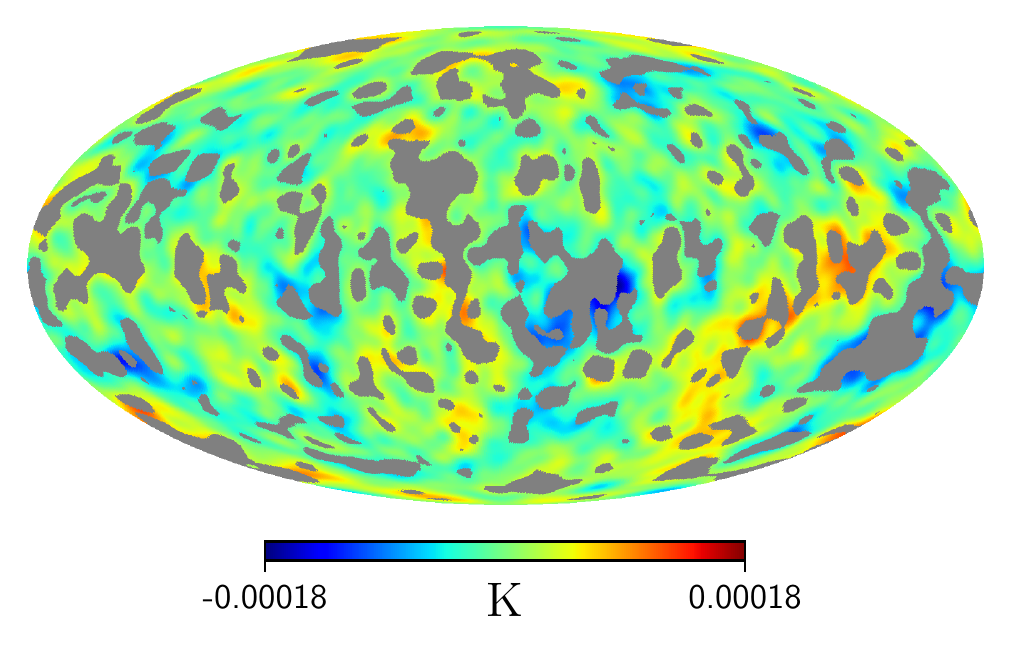}
    \includegraphics[width=0.45\textwidth]{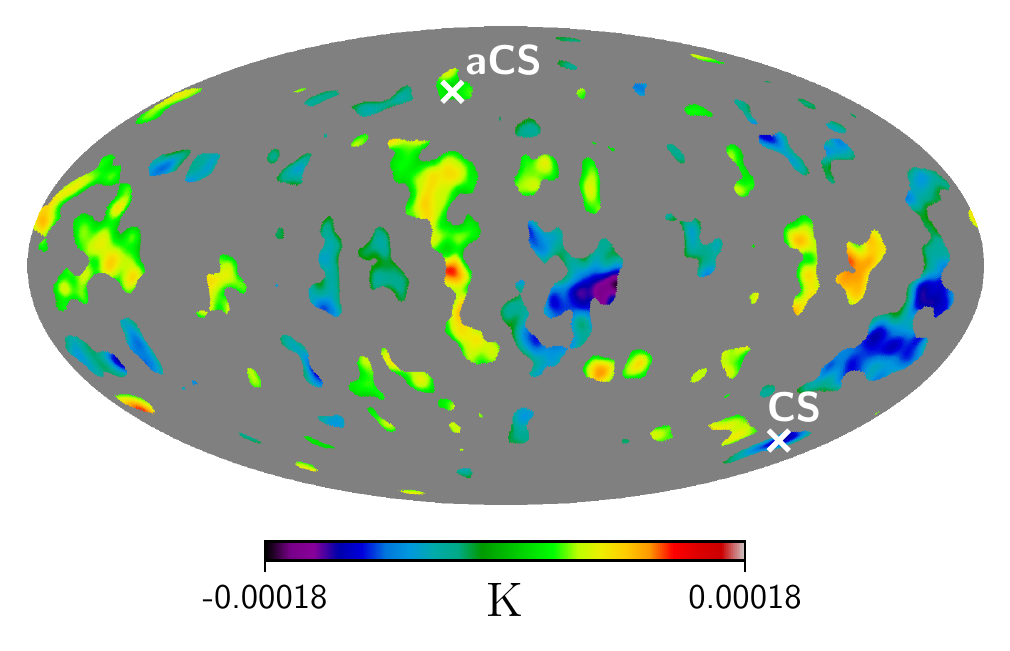}      \includegraphics[width=0.45\textwidth]{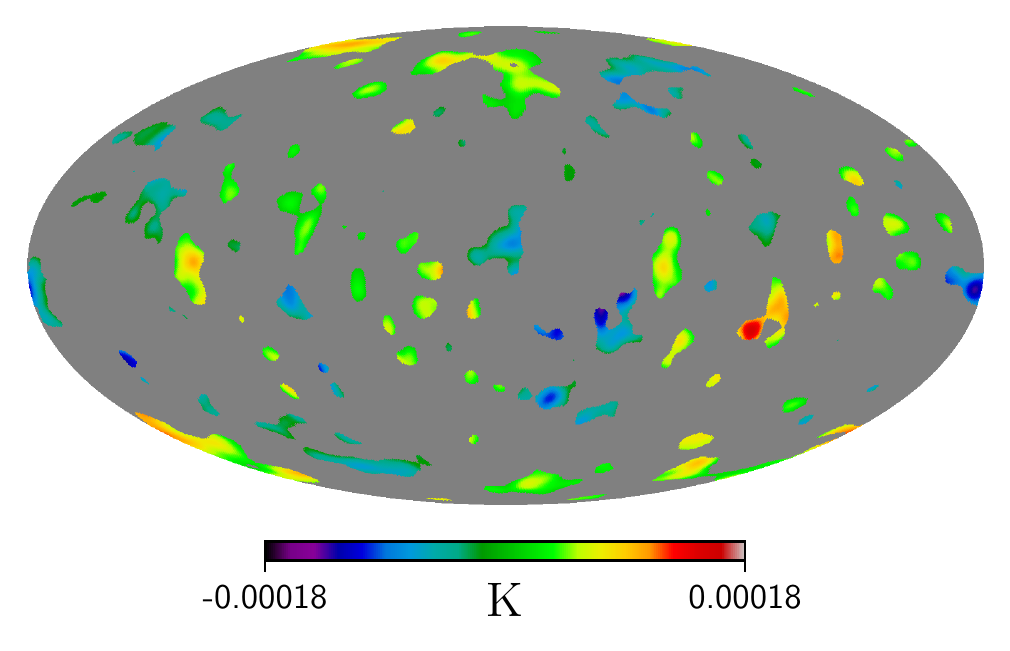} 
    \caption{Top left: The zones of the sky that are located in the left tail of the distribution function are shown in blue (regions of asymmetric excess), and the zones of the sky from the right tail are shown in red (symmetric excess). Top right: the zones of the SMICA
    map with $Z$ density $D>10^{8}\,\mathrm{K}^{-2}$.  Bottom left: the zones
    of SMICA map with density $D<10^{8}\,\mathrm{K}^{-2}$. Bottom right: the same as left, but for even $\ell$-excess. The labels CS and aCS mark the WMAP/Planck Cold Spot. }
    \label{fig:zones}
\end{figure}

To test the significance of departures from the Gaussian model, we run simulations based on a power spectrum from CAMB \cite{Lewis_2000} with Planck 2018 best fit cosmological parameters from \cite{Planck_2020, PS2018}. This is shown in figure~\ref{fig:hists}.
The main result is a $\sim3\sigma$ deficit on the right (symmetric, even-$\ell$) side of the distribution function. 
Also, in figure~\ref{fig:zones}, we show the parts of the sky contributing to the tails of the distribution function with density $D\le 10^{8}\,\mathrm{K}^{-2}$. 
To show the morphology of asymmetry, we use the SMICA map as a background and masked out all zones which contributed to symmetric and asymmetric modes (see the four maps in figure~\ref{fig:zones}). The upper right map corresponds to SMICA map with $\Theta=5^\circ$ with the symmetric and asymetric zones with $D<10^{8}\,\mathrm{K}^{-2}$ masked out. The bottom maps correspond to the SMICA signal for the asymmetric tail (on the left) and the symmetric tail (on the right).

It is important to note that all known local anomalies of the CMB sky contribute to the asymmetric part. The peaks 1a, 1b, 2a, and 2b, discussed above, the Galactic Cold Spot, the WMAP/Planck Cold Spot and its counterpart
(see figure~\ref{fig:zones}) are  presented there. At the same time, from figure~\ref{fig:hists}, the odd-$\ell$ asymmetric modes lie within the range of Gaussian simulations and show no $Z$-distribution anomalies when $\Theta=5^\circ$. In contract, the even-$\ell$ modes reveal significant departure from Gaussian simulation (greater then $99.7\%$), and they are free
from the contribution of  the local anomalies listed above. This result is in agreement with \cite{Kim_2010,Kim_2012}, where it was pointed out that odd-$\ell$ tail of the WMAP power spectrum perfectly follows to the best fit $\Lambda$CDM model, while the even-$\ell$ tail significantly departs from it. 

The analysis presented above of the parity asymmetry for Planck component separation products has a strong dependence on
the multipole domain under consideration. The maximum of asymmetry corresponds to multipoles $20\le\ell\le 30$. In our pixel domain approach that means that the significance of asymmetry should depend on the smoothing angle $\Theta$. This statement is illustrated in the right panel of figure~\ref{fig:hists}, where the smoothing angle is $\Theta=2.5^\circ$ and the range of
the corresponding multipoles is extended up to $40$--$50$. Here the asymmetric tail has no anomalies, while the symmetric tail is still marginally peculiar for high positive $Z$. From figure~\ref{fig:zones} (the bottom left and right panels) one can see the reason for that: in the symmetric component
we have a deficit of the negative peaks in the comparison with the asymmetric component. 

We have focused the analysis on the tails of the $Z(\vb{n})$ distribution function, corresponding to the highly symmetric and asymmetric peaks.
Although it is not visible in figure~\ref{fig:hists} due to the logarithmic axes scale in use, there is also a slight excess of low-$Z(\vb{n})$ pixels, complementary to the general deficits found in the tails, as constrained by the histogram estimator.

\section{\boldmath Are Gaussian simulations even-\texorpdfstring{$\ell$}{l} dominated?}
\label{sec:5}

One of the features of the best fit $\Lambda$CDM cosmological model based on the Planck 2018 power spectrum  is the even-$\ell$ parity dominance (on average), at least for the multipole domain $\ell \le 50$. This fact attracted less attention in the literature devoted to investigation of parity anomaly, but it is very important for understanding its origin. Actually,
the even-$\ell$ dominance can be quantitatively seen right from the CMB power spectrum, which is close to $C_\ell \propto \ell(\ell+1)$ for $2\le\ell\le 50$ \cite{Kim_2010}. For a power spectrum of this form, the highest amplitude corresponds to the quadrupole and it will determine the even-$\ell$ asymmetry of $Z(\vb{n})$.  For the actual Planck 2018 power spectrum we 
confirmed that result by simulations presented in figure~\ref{fig:hists}. Here one can see that the mean value of $Z(\vb{n})$ over all realisations and the sample variance (blue lines and region) are skewed towards the symmetric components, while the actual SMICA data reveals the odd-$\ell$ asymmetry. 

One of the key reason for dominance of the asymmetric modes is related to the
abnormally low quadrupole.
The analysis of the quadrupole anomaly and its influence on the distribution of $Z(\vb{n})$ is presented below in two directions.

\begin{figure}[!t]
    \centering
    \includegraphics[width=0.48\textwidth]{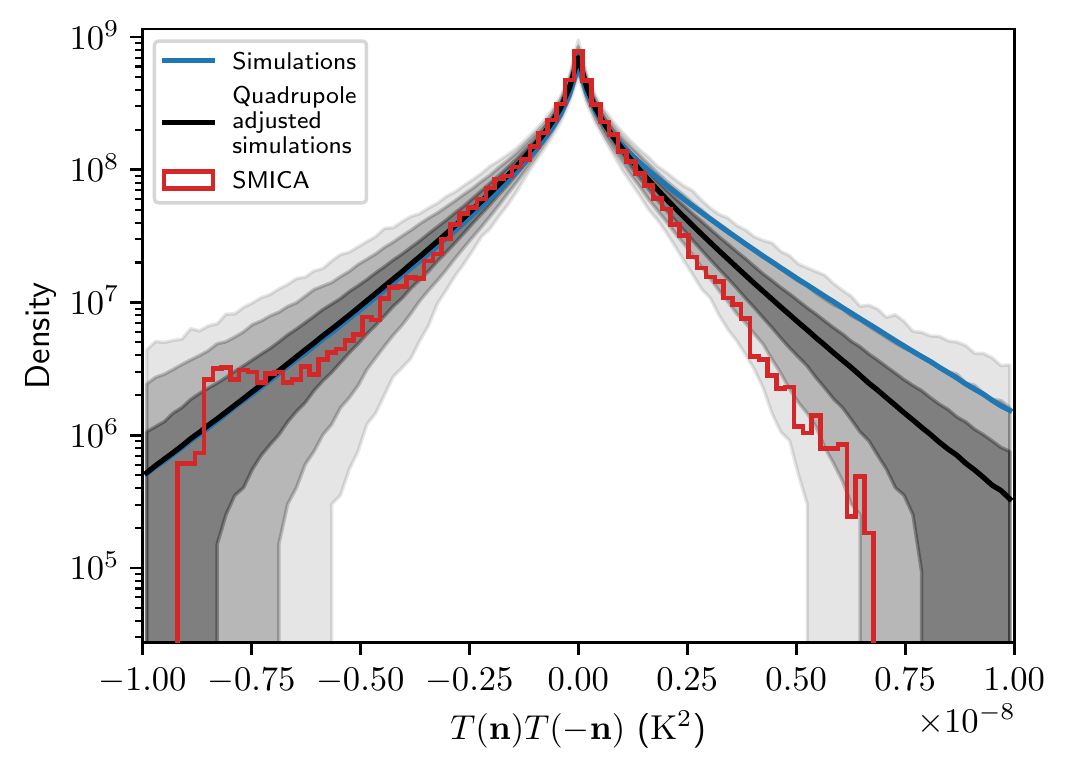}
    \includegraphics[width=0.48\textwidth]{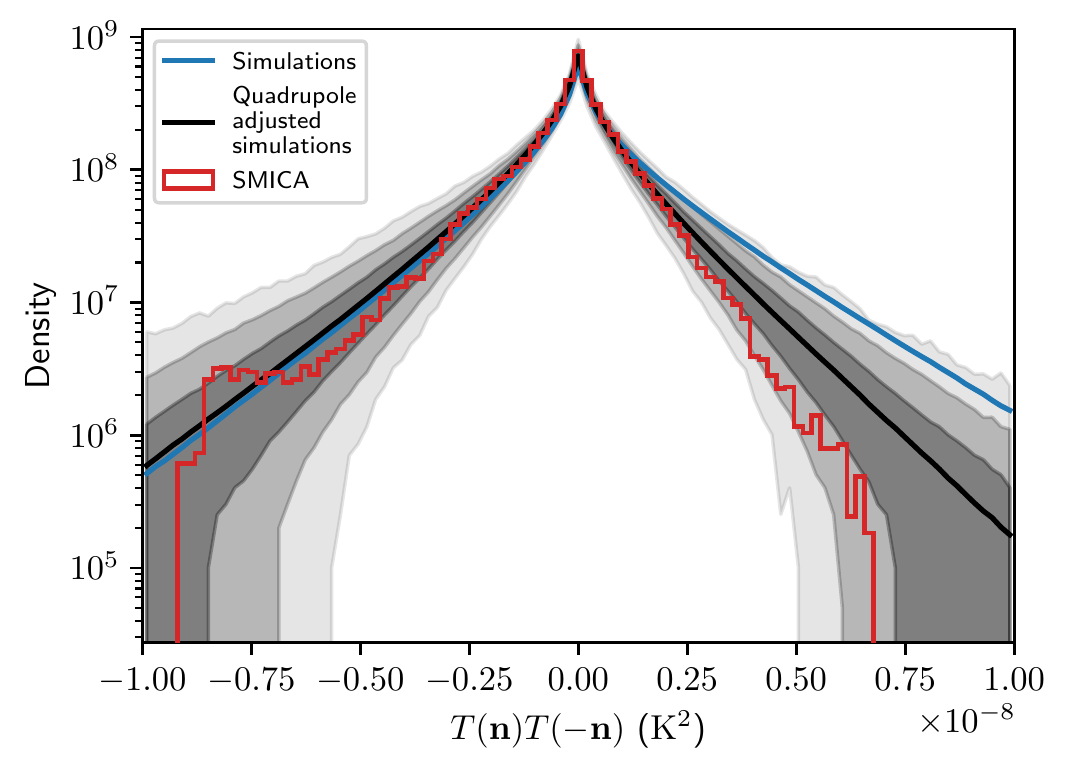}
    \caption{Two different methods of simulating quadrupole-adjusted maps. In the left panel, the $\ell = 2$ coefficient in the best-fit power spectrum is adjusted to that of SMICA, and simulations are generated from this modified power spectrum. In the right panel,  simulations are generated from the original power spectrum, and each map is manually adjusted to the SMICA quadrupole. The black shaded regions show the 68\%, 95\%, and 99.7\% uncertainty of the density estimation. Comparison of the two figure shows that the two methods of simulation give almost the same results. However, the variance for the power spectrum adjustment (left panel) is slightly higher than for the map adjustment (right panel). The power spectrum adjustment is the standard method used in this work.}
    \label{fig:quadrupole_adjustment}
\end{figure}

First, we generate a power spectrum from CAMB using the best-fit cosmological parameters from Planck, and calculate realizations of $Z(\vb{n})$ from simulated Gaussian maps from this power spectrum.\footnote{In this article we will use Gaussian simulations from HEALPIX \texttt{synfast} program. The comparison of them and Planck FFP simulations can be found in \cite{Muir_2018}.}  The average histogram of $Z(\vb{n})$ and the variation in the counts of the histograms was shown in figure~\ref{fig:hists} by the blue shaded regions.
Alternatively, the second method of simulations is based on a new power spectrum in which the $\ell = 2$ amplitude has been reduced to the SMICA quadrupole power. This is a very large reduction, because the SMICA quadrupole is abnormally small. From this new quadrupole-adjusted power spectrum, simulations are generated and the histograms are calculated like before. The results are shown by the black solid line and black shaded regions in figure~\ref{fig:quadrupole_adjustment} (left panel). As it is seen from that figure, the significance level of abnormality for the even-$\ell$ multipoles drops from about $3\sigma$ to $2\sigma$. One can treat this result as a formal solution of the problem of parity asymmetry, but the price of it is very high.  Formally, we move out from the best-fit $\Lambda$CDM cosmological model with statisticaly abnormal quadrupole to the models with normal low quadrupole, but more complicated physics of the CMB anisotropy in the Sachs-Wolfe domain.\footnote{Note that at the level of $2\sigma$ we still have symmetric modes anomaly for $Z(\vb{n})$ after modification of the quadrupole.}
The third method of simulations is based on the Planck best fit cosmological model, as in figure~\ref{fig:hists}, but for each realisation of the CMB map the quadrupole component  is set equal to the SMICA quadrupole.  With this brute-force approach, we completely
remove the effect of sample variance for the quadrupole, which should be viewed as a kind of artificial model of systematic effects. The difference is slight, but as one can see from the right panel of figure~\ref{fig:quadrupole_adjustment},
the significance of parity anomaly of symmetric modes is reduced almost to the $68\%$ confidence level.

\subsection{Variation of cosmological parameters}

Discussion of the role of the quadrupole in the framework of the modified Planck 2018 cosmological model raises the more general question of how much the statistical significance of anomalies depends on cosmological parameters.
In answering this question, we looked at two artificial models, changing the cosmological parameters manually. The first modification consists in replacing the Hubble constant $67.5$ to $71.0$ km/s/Mpc without changing the other parameters. The second model was discussed in \cite{Kim_2012} in relation to fitting the power spectrum of even multipoles, in which the main factor is the use of the Harrison-Zeldovich spectral index. 
Hereafter we will call that model as the ``even model''.
In the left column of table~\ref{tab:parameters} is shown the values of the cosmological parameters under consideration, as derived from the Planck 2018 combined analysis \cite{Planck_2020} with the corresponding error bars at 68$\%$ confidence level. 
The simulations used so far in this work are consistent with these parameters.
The right column of table~\ref{tab:parameters} shows alternative parameters for the even model.

\begin{table}[h!]
    \centering
    \begin{tabular}{|r|l|l|}
        \hline
        Parameter & Planck 2018 \cite{Planck_2020} & Even model \cite{Kim_2012} \\
        \hline
        \hline
        $H_0$ 
        &  $67.37 \pm 0.54$ & $71.0$ \\
        $\Omega_\mathrm{b} h^2$ &  $0.02233 \pm 0.00015$ & $0.0235$  \\
        $\Omega_\mathrm{c} h^2$ &  $0.1198 \pm 0.0012$ & $0.109$ \\
        $\tau$ & $0.054 \pm 0.0074$ & $0.06$\\
        $n_\mathrm{s}$ & $0.9652\pm 0.0042$ & $0.995$\\
        \hline
    \end{tabular}
    \caption{Cosmological parameters derived in the Planck 2018 analysis \cite{Planck_2020} compared to even model.}
    \label{tab:parameters}
\end{table}

\begin{figure}[t]
    \centering
    \includegraphics[width=0.39\textwidth]{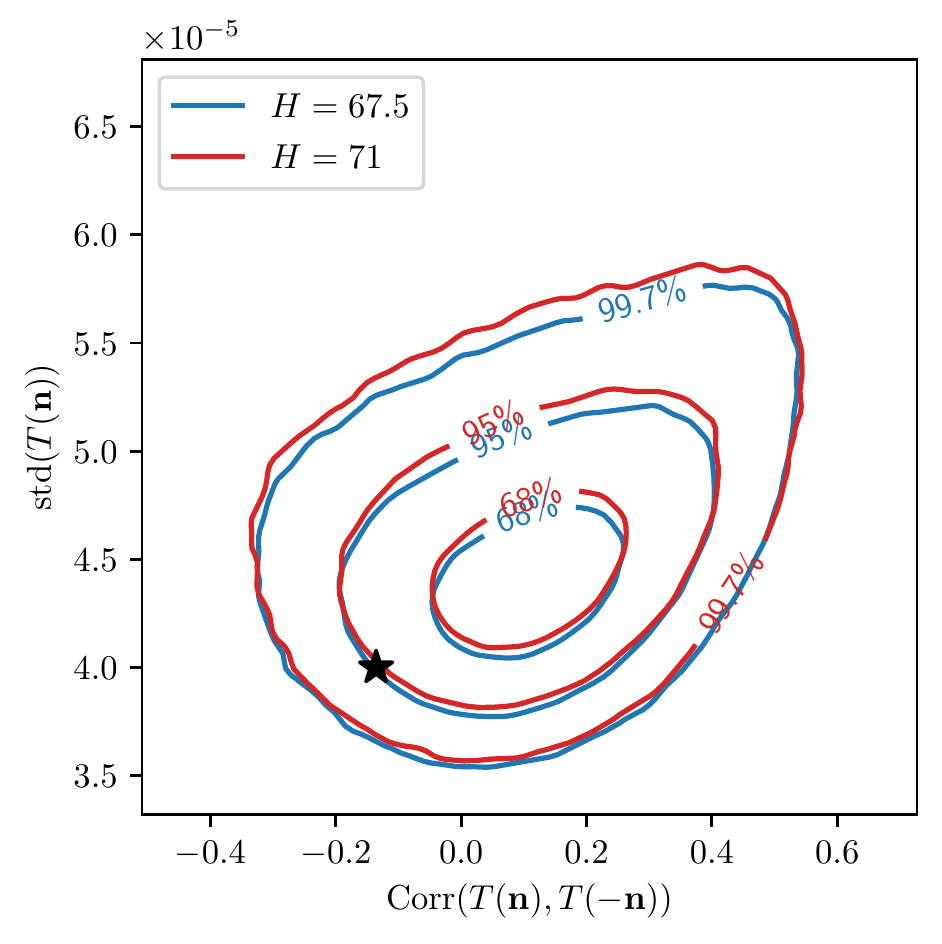}
    \caption{Contours of $\mathrm{Corr}(T(\vb{n}), T(-\vb{n}))$ and $\mathrm{std}(T(\vb{n}))$ within $\Lambda$CDM models having $H_0=67.5$ (blue) and $H_0 = 71.0$ (red).}
    \label{fig:contour}
\end{figure}

In the Gaussian theory, the parameters $\rho = \mathrm{Corr}(T(\vb{n}), T(-\vb{n}))$ (see  equation~(\ref{eq:rho}))  and $\mathrm{std}(T(\vb{n}))$ determine the shape of the distribution function of $Z(\vb{n})$.
Both parameters show dependence on the Hubble parameter.
Using the 2-sample Kolmogorov-Smirnov test, the difference in $\rho$ is detectable with p-value $<10^{-5}$ between the simulations from $H=67.5$ and the simulations from $H=71$. $\mathrm{std}(T(\vb{n}))$ is also detectable with p-value $< 10^{-5}$ using the same test between the models for $H=67.5$ and $H=71$. 
The contours for these parameters are shown in figure~\ref{fig:contour}.

\begin{figure}[t]
    \centering
    \includegraphics[width=0.325\textwidth]{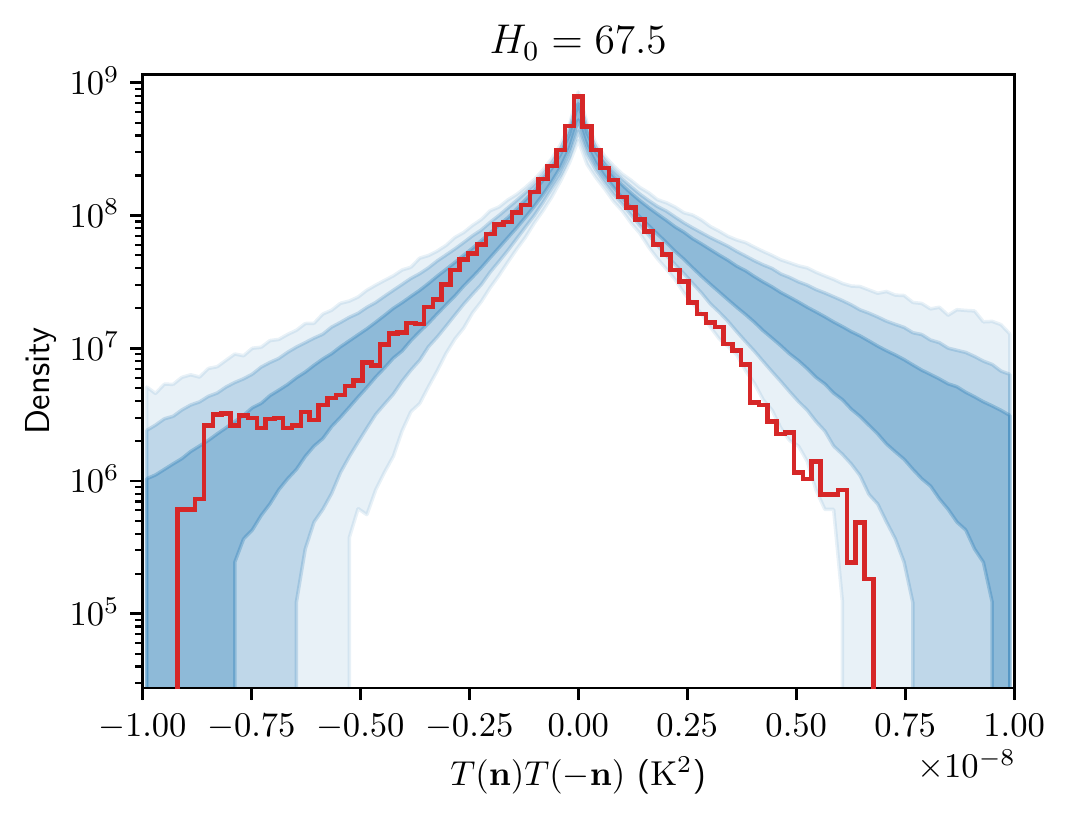}
    \includegraphics[width=0.325\textwidth]{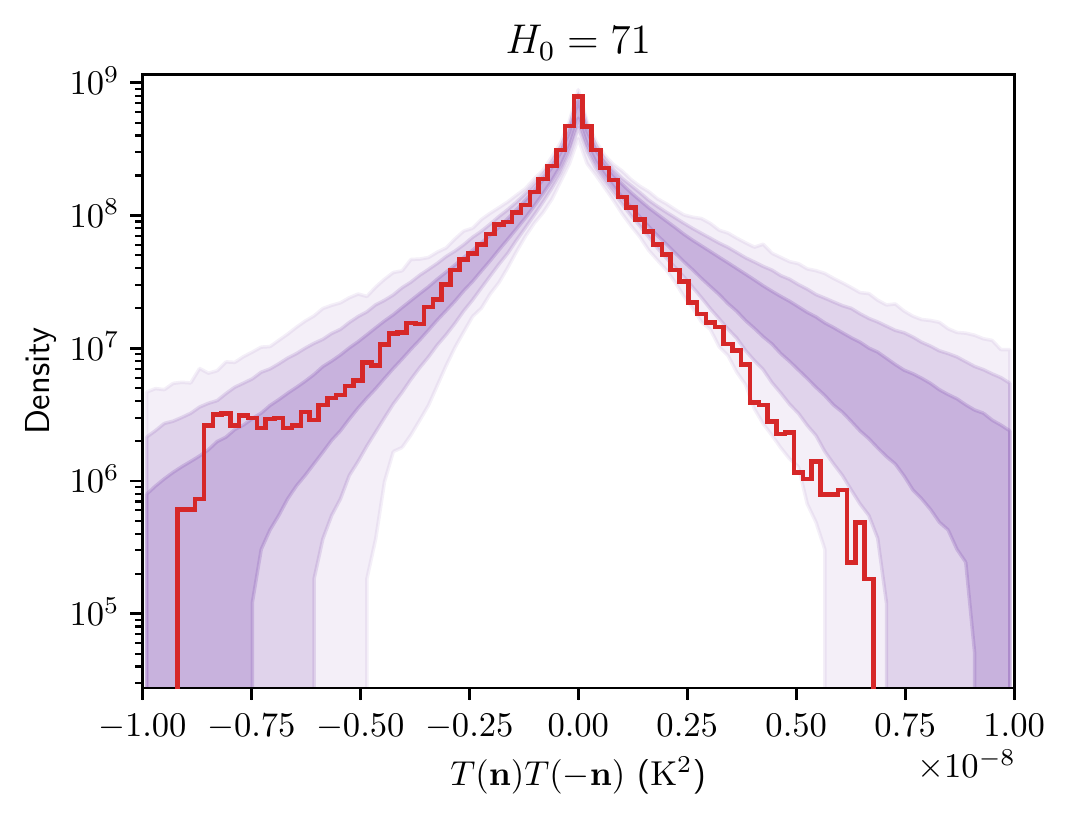}
    \includegraphics[width=0.325\textwidth]{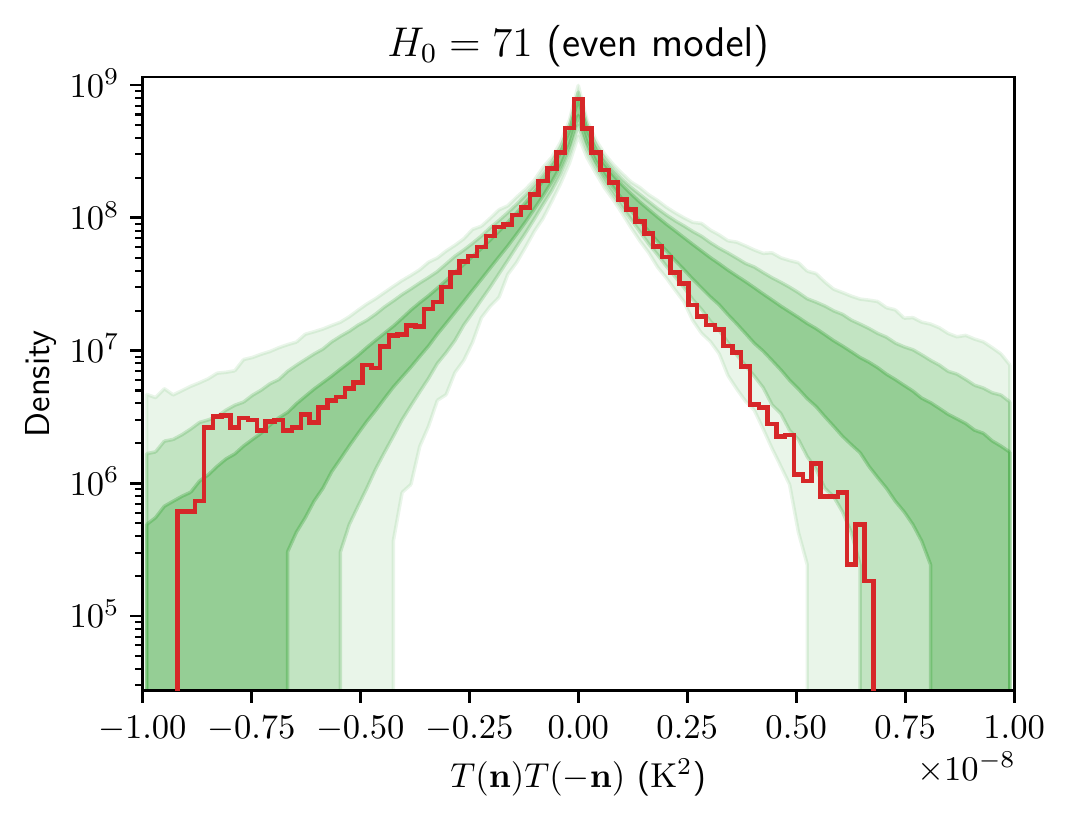}
    \caption{From left to right: normal model, normal model with $H_0$ adjusted to $71$, and even-model (see table~\ref{tab:parameters}). The most important observation is the significance of the bump on the left side of the distribution function, which contains the peaks 1a/1b/2a/2b and other high amplitude asymmetric peaks discussed in Section \ref{sec:3} and shown in figure~\ref{fig:map}.}
    \label{fig:parameters}
\end{figure}

In figure~\ref{fig:parameters}, the range of simulations is shown for three models: default model with the best-fit Planck parameters (this is effectively the same as the left panel of figure~\ref{fig:hists}), the second  model derived by increasing the Hubble parameter from $67.5$ to $71.0$ while leaving all other parameters fixed, and third the even-model from the right column of table~\ref{tab:parameters}.
The general trend is a  shift of density out of the tails. As we advance through these models, the significance of the even-$\ell$ deficit of SMICA weakens from around $3\sigma$ to around $2\sigma$; correspondingly, the peak odd-$\ell$ excess becomes a little bit more significant, nearing a maximum departure of $2\sigma$ for the even-model.

\section{Power and parity asymmetries}
\label{sec:6}

The power symmetry is one of the most intriguing CMB anomalies. It leads to weak coupling between neighbouring multipoles and an anisotropic distribution of the power spectrum. For the model of power asymmetry based on dipole modulations, the observed CMB temperature $T(\vb{n})$ is related with statistically isotropic Gaussian component $T_g(\vb{n})$ as:
\begin{equation}
T(\vb{n})=\left(1+(\vb{m}\vdot\vb{n}) D\right)T_g(\vb{n})
\label{dipole}
\end{equation}
where $\vb{m}$ is unit vector in the direction of dipole, $\vb{m}\vdot\vb{n}$ denotes the dot product, and $D$ is the amplitude of the dipole modulation. 
The dipole modulation of the statistically isotropic Gaussian random signal $T_g(\vb{n})$ means that
in the direction of the maximum of the dot product $\vb{m}\vdot\vb{n}$ we have
amplification of the local amplitude of the signal, while in opposite direction, it is
effectively decreasing. 

For characterisation of the asymmetry induced by the dipole modulation of
the isotropic Gaussian field $T_g(\vb{n})$, we use the following estimator:
\begin{equation}
G(\vb{n})=T^2(\vb{n})-T^2(-\vb{n})=4S(\vb{n})A(\vb{n}).
\label{pow1}
\end{equation}
Using equation~(\ref{dipole}), this estimator can be presented in the following form: 
\begin{equation}
G(\vb{n})=\left[-R(\vb{n})+2D\cos(\xi)(2+R(\vb{n}))-R(\vb{n})D^2\cos^2(\xi)\right]T^2_g(\vb{n})
\label{pow2}
\end{equation}
where $\xi$ is the angle between $\vb{m}$ and $\vb{n}$, and $R({\vb{n})+1=\frac{T^2_g(\vb{-n})}{T^2_g(\vb{n})} }$ is the asymmetry parameter. If $R(\vb{n})\ll D$ we get:
\begin{equation}
G(\vb{n})\propto S(\vb{n})A(\vb{n})\simeq D\cos(\xi)T^2_g(\vb{n}).
\label{pow3}
\end{equation}

\begin{figure}[!t]
   \centering
    \includegraphics[width=0.8\textwidth]{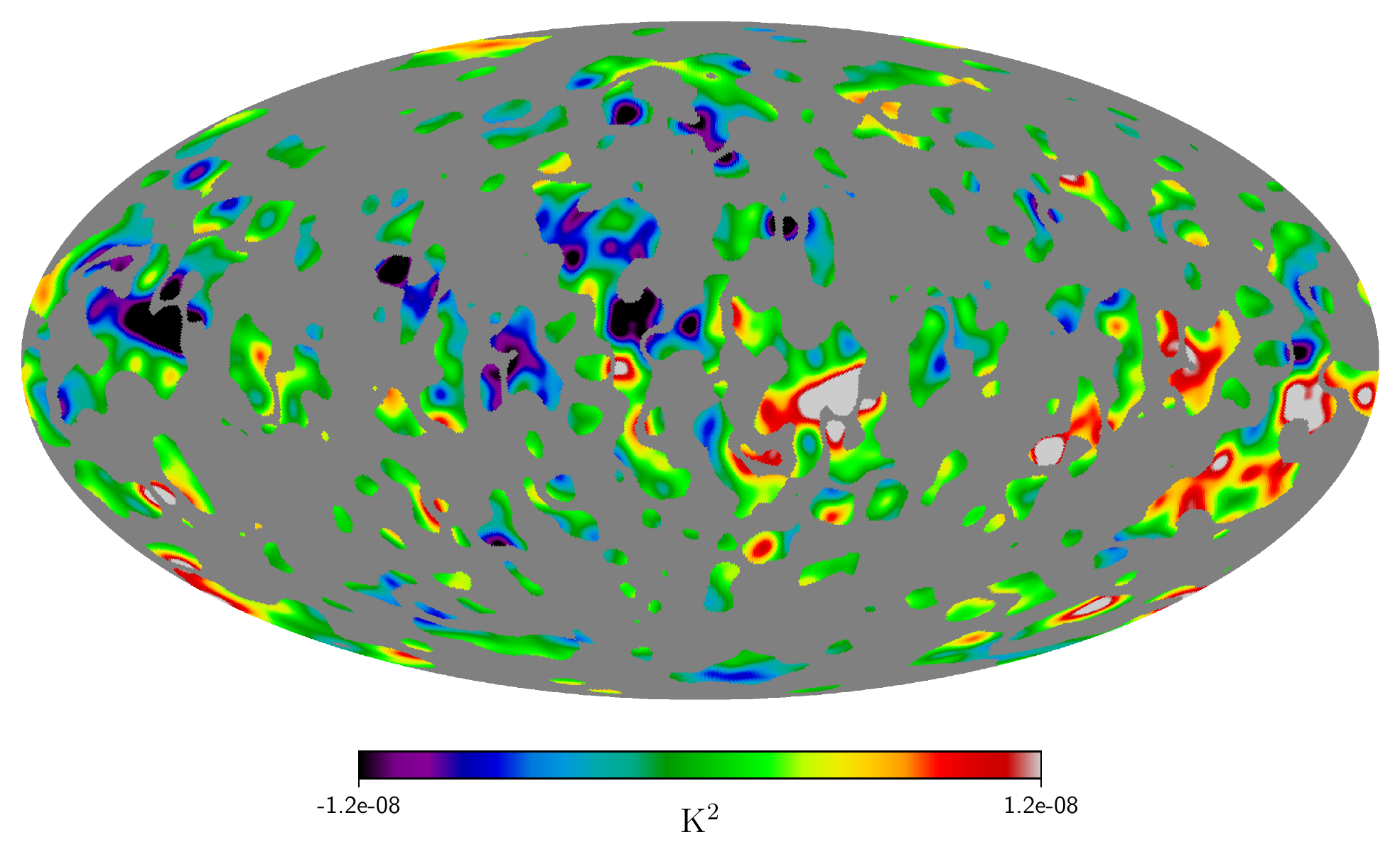}
    \caption{$G(\vb{n})$ in zones where the density of $Z(\vb{n})$ is $\le 10^8\,\mathrm{K}^{-2}$ for the SMICA map.}
    \label{fig:power}
\end{figure}
Thus, from equation~(\ref{pow3}) one can see that combination $S(\vb{n})A(\vb{n})$ of the symmetric and asymmetric components of the observable signal can be used as an estimator of the dipole modulation from equation~(\ref{dipole}). For illustration of this statement in figure~\ref{fig:power} we show 
the map $S(\vb{n})A(\vb{n})$ from  SMICA signal in zones with $Z$-density $\le 10^8 \,\mathrm{K}^{-2}$
in order to get the parity asymmetry and the dipole modulation simultaneously.

\begin{figure}[!t]
   \centering
    \includegraphics[width=0.48\textwidth]{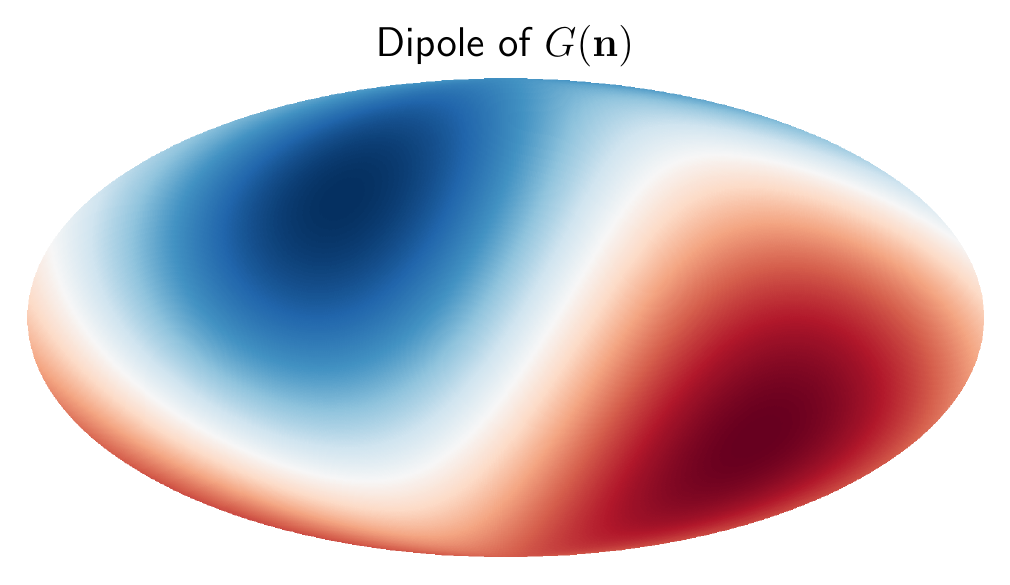}
    \includegraphics[width=0.48\textwidth]{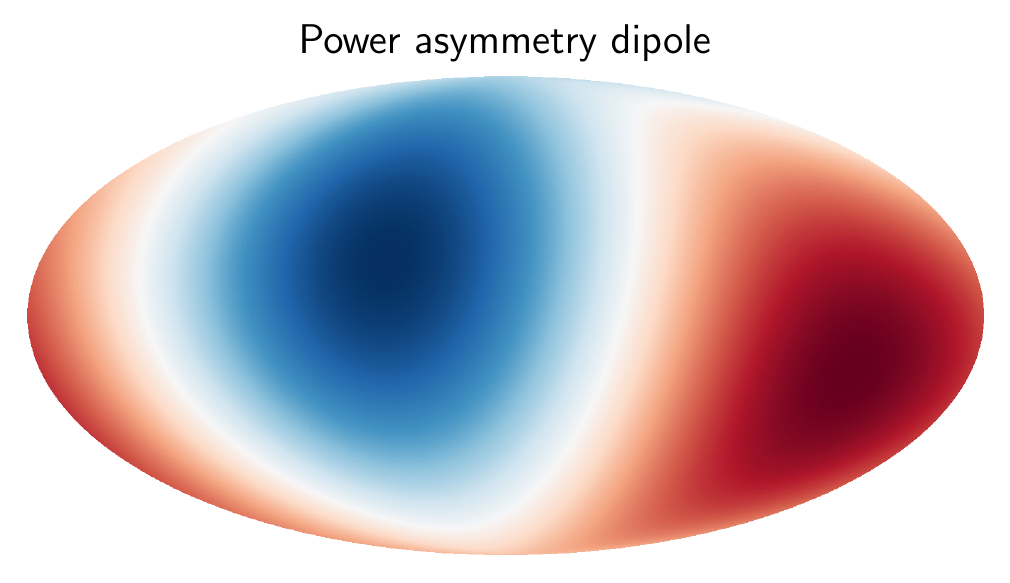}
    \caption{The $G(\vb{n})$ dipole for the SMICA map and dipole of modulation from equation~(\ref{dipole}).}
    \label{fig:power-dipole}
\end{figure}

\begin{figure}[!t]
   \centering
    \includegraphics[width=0.48\textwidth]{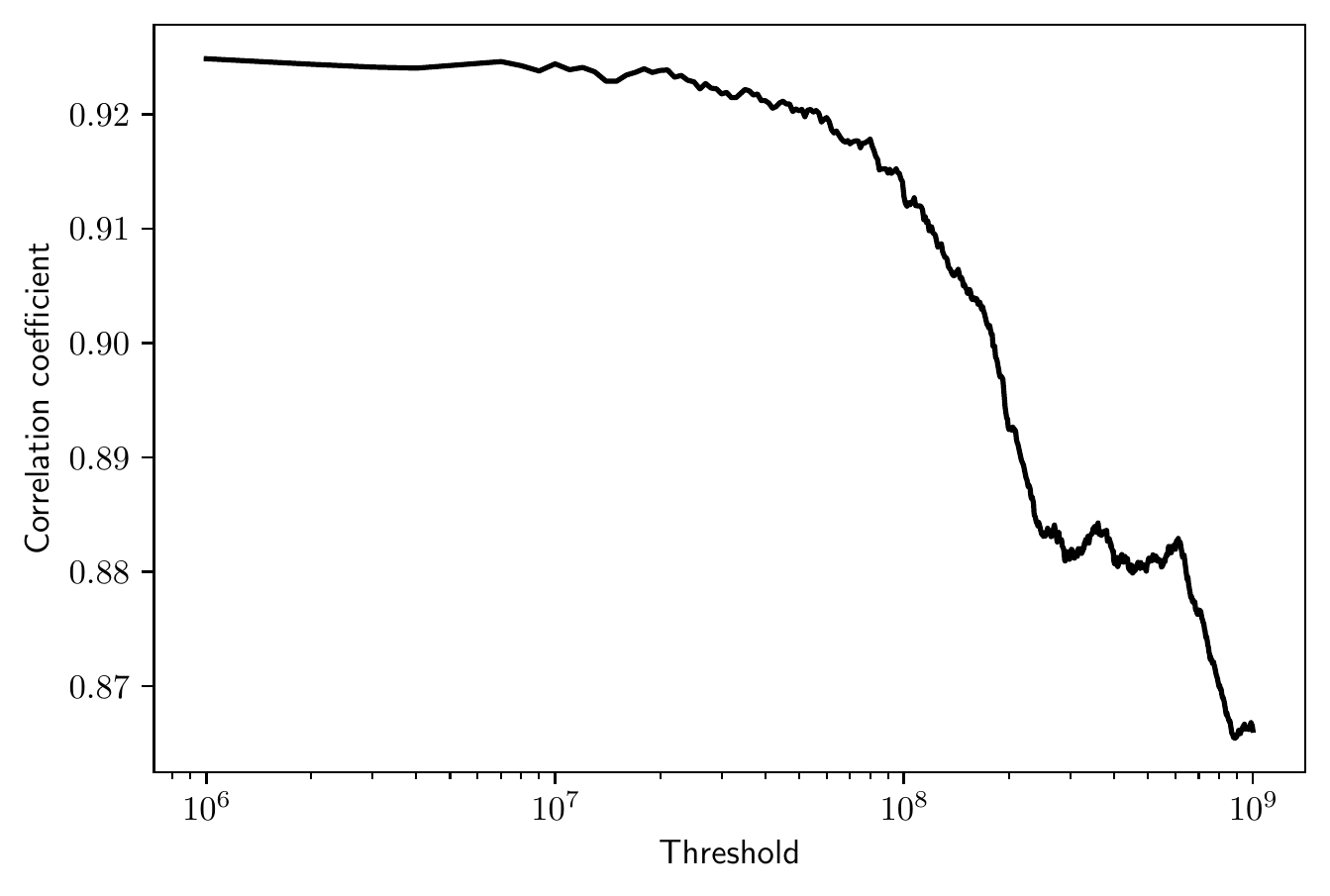}
    \caption{Correlation coefficient between dipole maps as a function of the threshold of the $Z(\vb{n})$ density below which the correlation is computed.}
    \label{fig:threshold}
\end{figure}

We decompose this map into spherical harmonics and get a dipole
component shown in figure~\ref{fig:power-dipole}. For comparison, in this figure, we constructed a dipole  power modulation model (see equation~(\ref{dipole})) and estimated the pixel-to-pixel Pearson cross-correlation coefficient, which is $ C = 0.87 $.
This coefficient is a function of the threshold amplitude $Z(\vb{n})$. 
In figure~\ref{fig:threshold} we plot the dependency of the dipole-dipole cross-correlation versus the amplitude of the threshold.
The cross-correlation coefficient is only computed within the zones where the estimated density of $Z(\vb{n})$ is below the threshold, and then we repeat this for different thresholds. The high threshold corresponds to using more of the sky for the $C$: then we get 0.87. 
Lower thresholds correspond to calculating $ C $ only in areas with the most asymmetric pixels: then we get convergence to $ C = 0.925 $. Asymptotically, it will be associated only with the peaks 1a/1b and 2a/2b discussed in Section \ref{sec:3}.

\section{Conclusion}
\label{sec:7}

In the pixel domain, we investigated the sources of asymmetry hovering in the CMB temperature maps. We have demonstrated that this type of anomaly is mainly associated with the presence of four high amplitude positive and negative peaks (1a, 1b, 2a, 2b), symmetrically spaced from each other relative to the Galactic center.
Peaks 1a and 1b are associated with the zones belonging to the Northern Galactic Spur and the direction of the dipole modulation of the power spectrum
CMB anisotropy. 
The influence of the Northen Galactic Spur on the CMB signal has been
pointed out in \cite{Liu_2014}. The two other peaks (2a, 2b) are associated with the Galactic plane (the Galactic Cold Spot and its partner asymmetric in amplitude).
The same type of asymmetric peaks, but with a smaller amplitude, belongs to
WMAP/Planck Cold Spot and its partner in the Northern Galactic Spur.
The most striking result of our analysis is the fact that
these local anomalies of the CMB map increase the level of the asymmetric part
the distribution of the estimator $Z(\vb{n})$ to a level consistent with predictions based on simulations of a random Gaussian process.
In contrast, the deficit of peaks symmetric in amplitude and position is accompanied by a decrease in the symmetric part $Z(\vb{n})$ of the $3\sigma$ level, which is the source of the parity asymmetry of the CMB temperature maps. 

Another source of the deficit of the symmetric part of the $Z(\vb{n})$ function is the low amplitude of the SMICA quadrupole. We investigated the influence of this factor using Gaussian simulations, in which, instead of the theoretical value in the power spectrum, we used the amplitude of the SMICA quadrupole. Moreover, this modified power spectrum still corresponds to the Planck 2018 best fit cosmological model, with the exception of the quadrupole. We have shown that such a modification of the power spectrum leads to a weakening of the confidence level of $Z(\vb{n})$ for symmetric modes in simulations and reduces the significance of the parity asymmetry to the $2\sigma$ level.

The cosmological parameters are also reflected in the distribution function of $Z(\vb{n})$ 
and their effect on 
the significance of anomalies. We have considered alternative models in which the Hubble parameter is increased to $H_0 = 71.0$ from the Planck best-fit value of around $67.5$. The model with higher $H_0$ has a slightly more symmetric distribution function. Further adjustment of the cosmological parameters to the ``even model'' derived from \cite{Kim_2012} has the same overall effect on the $Z(\vb{n})$ function. 

We investigated the relationship between the asymmetry of the power spectrum and the level of the parity asymmetry  in the framework of a model with dipole modulation of a statistically uniform Gaussian signal. We have shown that the dipole modulation amplitude is directly related to the level of asymmetry of the $G(\vb{n})=S(\vb{n})A(\vb{n})$ estimator, which is the product of the symmetric and asymmetric components of SMICA in the zones of
maximum asymmetry of the function $Z(\vb{n})$.  The dipole
the component of the $G(\vb{n})$ estimator in the  zones with different levels of parity asymmetry  practically coincides with  the dipole modulation of the power spectrum: the corresponding Pearson cross-correlation coefficient $C$ is localised in the domain  $0.87-0.92$. Again, the maximum of $C=0.92$ achieved from the peaks 1a, 1b, 2a, 2b, where
the parity asymmetry has a point of maxima.

In conclusion, we would like to note the peculiarity of our analysis of various anomalies in the pixel domain. At first glance, it may seem trivial that various anomalies turned out to be interdependent due to their
joint contribution to each pixel. However, one must not forget that the transition from
pixel to multipole analysis of some anomalies is associated with integration over the map, in which a significant role belongs to different types of masks with different sky coverage. Thus, the zones identified by us in the map must be taken into account (masked out) in
analysis of anomalies in a multipole domain.

\appendix
\acknowledgments

The  HEALPix  pixelization  \cite{Gorski_2005}  scheme  was  heavily used in this work, and we thank them for their contributions to the field.  This research was partially funded by  Villum Fonden through the Deep Space project.

\bibliography{references.bib}

\end{document}